\documentclass[a4paper,11pt]{article}
\usepackage{pos}

\usepackage{epsfig}
\usepackage{graphicx}
\usepackage{palatino}
\usepackage[english]{babel}
\usepackage{hyphenat}

\usepackage{mathtools}
\usepackage{mathrsfs}
\usepackage{bbm} 
\usepackage{bm}
\usepackage{slashed}
\usepackage{epstopdf}
\usepackage{xcolor}
\usepackage{booktabs}
\usepackage{float}
\definecolor{lcolor}{rgb}{0.,0.0,0.}
\definecolor{citcolor}{rgb}{0,0.,0.5}

\def\be{\begin{eqnarray*}}
\def\ee{\end{eqnarray*}}
\def\beq{\begin{eqnarray}}
\def\eeq{\end{eqnarray}}

\newcommand{\bea}{\beq \begin{aligned}}
\newcommand{\eea}{\end{aligned}\eeq}

\newcommand{\br}{{\boldsymbol r}}
\newcommand{\bx}{{\boldsymbol x}}

\newcommand{\rme}{{\rm e}}
\newcommand{\rmd}{{\rm d}}

\def\abar{\bar\alpha_s}

\title{Scheme Transformations in Non-Linear QCD Evolution: Towards a Stable NLO BK Equation}
 \ShortTitle{Scheme transformations in non-linear QCD evolution: towards a stable NLO BK equation}

\author[a]{Renaud Boussarie}
\author[b]{Paul Caucal}
\author*[c]{Piotr Korcyl}
\author[d]{Yacine Mehtar-Tani}

\affiliation[a]{CPHT, CNRS, \'Ecole polytechnique, Institut Polytechnique de Paris, 91128 Palaiseau, France}
\affiliation[b]{SUBATECH UMR 6457 (IMT Atlantique, Universit\'{e} de Nantes,
IN2P3/CNRS), 4 rue Alfred Kastler, 44307 Nantes, France}
\affiliation[c]{Institute of Theoretical Physics, Jagiellonian University, ul. Lojasiewicza 11, 30-348 Krak\'{o}w, Poland}

\affiliation[d]{Physics Department, Brookhaven National Laboratory, Upton, NY 11973, USA}

\emailAdd{renaud.boussarie@polytechnique.edu}
\emailAdd{caucal@subatech.in2p3.fr}
\emailAdd{piotr.korcyl@uj.edu.pl}
\emailAdd{mehtartani@bnl.gov}

\abstract{High-energy QCD evolution at NLO suffers from large collinear logarithms that make the perturbative expansion of the Balitsky-Kovchegov (BK) equation unstable. We develop a systematic scheme-transformation framework using the Doi-Peliti formalism to relate projectile ($k^+$) and target ($k^-$) factorization schemes. The transformation of high-energy operators and the evolution kernel generates NLO terms that cancel the problematic double collinear logarithm. The resulting NLO BK equation is free of this instability, with only single collinear and double anti-collinear enhancements remaining. Numerical solutions show stable evolution up to $Y=10$ for both Golec-Biernat-Wusthoff and McLerran-Venugopalan initial conditions, providing a consistent path toward combining stable NLO evolution with rotated NLO impact factors.
}

\FullConference{
}

\begin{document}
\maketitle

\section{Introduction}

Over the past several years, significant progress has been made in computing small-$x$ observables at next-to-leading order within the framework of high-energy factorization. These developments include inclusive deep inelastic scattering (DIS), forward hadron production, inclusive dijet production, and semi-inclusive DIS (SIDIS). At the same time, this higher-order program has encountered several conceptual and phenomenological challenges. These include the instability~\cite{Lappi:2015fma} of the NLO BK evolution equation~\cite{Balitsky:2007feb,Balitsky:2009xg}, the appearance of negative cross sections in forward hadron production~\cite{Stasto:2013cha}, and, more recently, Sudakov double logarithms with the opposite sign in less inclusive observables such as dijet production~\cite{Taels:2022tza,Caucal:2022ulg}.

The first efforts to address these issues focused on the NLO BK equation, where it was soon recognized that the instability originates from large collinear double and single logarithms. The double logarithms arise because the phase-space integration over the emitted gluon extends into a kinematically forbidden region, corresponding to $k^->P^-$, where $P^-$ denotes the dominant light-cone momentum of the target proton in DIS. Physically, this region corresponds to gluon fluctuations with formation times shorter than the longitudinal extent of the target shock wave and therefore lies outside the domain of validity of the high-energy operator product expansion. This mechanism is closely related to an earlier observation in the context of linear small-$x$ evolution~\cite{Salam:1999cn}, where the NLO BFKL kernel was found to be destabilized by large collinear contributions associated with higher-order poles in Mellin space.

In both the linear and nonlinear evolution equations, several remedies have been proposed. These include imposing kinematic constraints that enforce lifetime ordering of successive emissions, as well as explicit resummations of the collinear double logarithms~\cite{Beuf:2014uia,Iancu:2015vea}. In both cases, these strategies are implemented via a modification of the leading-order kernel of the BK equation. Besides these kinematically generated double logarithms, there also exist single collinear logarithms of physical origin, which are associated with the DGLAP anomalous dimension and therefore encode genuine collinear dynamics.

While these approaches substantially improve the numerical behavior of the evolution equations, they rely on resummations that are introduced in addition to the standard high-energy operator product expansion. As a result, they raise a number of conceptual questions regarding the underlying factorization scheme. In particular, it remains important to understand how these modifications affect the definition and universality of the high-energy operators, whether the factorization framework remains systematically under perturbative control, and how the evolution of the Color Glass Condensate operators is consistently matched to the hard coefficient functions entering physical observables. 

These considerations motivate the search for a formulation in which the collinear logarithms are systematically incorporated without modifying the operator content of the high-energy operator product expansion itself. Such a formulation would preserve the standard factorization structure while improving the perturbative behavior of the evolution order by order. In this paper, we present such a framework, based on nonlinear scheme transformations implemented at the level of physical observables within the high-energy product expansion, and demonstrate through a numerical study that the resulting NLO high-energy evolution equation is free of instabilities from double collinear logarithms.

\section{Doi-Peliti formalism for the Balitsky hierarchy}

We first briefly introduce the Doi–Peliti formalism,
 which allows us to formulate non-linear evolution, scheme transformations, and changes of basis for QCD operators in the saturation regime in a form reminiscent of linear transformations in standard linear algebra. In the large $N_c$ limit, any target operator $S$ can be written in term of the dipole operator $S_{12}$ in the fundamental representation, defined as
\begin{align}
    S_{12}\equiv \frac{1}{N_c}\textrm{Tr}(U_{\bx_1}U^\dagger_{\bx_2})\,,
\end{align}
where $U_{\bx_1}$ is an infinite light-like fundamental Wilson line at the fixed transverse coordinate $\bx_1$. One can then view a target operator as a vector $|S)$ in a Fock space spanned by the dipole-chain basis made of ordered strings of transverse coordinates: $|0) ,\;|\bx_1\bx_2),\; |\bx_1\bx_2\bx_3),\; |\bx_1\bx_2\bx_3\bx_4) \ldots \,,$ As such, $S^{(n)}=(\bx_{1}\bx_{2}\bx_{3}...\bx_{n}|S)$ represents $S_{12}S_{23}...S_{(n-1)n}$ and in the purely gluonic case, the dipole chain basis forms a complete set of states. Target operators involving fermion loops require to extend the dipole chain basis by including disconnected dipole chains represented as tensor products as in quantum mechanics, such as $|\bx_1\bx_4...)\otimes |\bx_3\bx_{2}...)$.

In the Doi-Peliti formalism, the Balisky hierarchy for the rapidity dependence of the target operators takes an exponential form
\begin{align}
    |S(\zeta))=\exp(\abar K(\abar) \ln(\zeta/\zeta_0))|S(\zeta_0))\,,
\end{align}
where $K=K_{\rm LO}+\bar\alpha_s K_{\rm NLO}+...$ is the operator generating the Balitsky hierarchy and $\bar\alpha_s=\alpha_s N_c/\pi$. Here $\zeta$ is defined as $\zeta =\rho^+P^-$ with $\rho^+$ the rapidity factorization scale along the positive light cone momentum direction (aligned with the dilute projectile) and $|S(\zeta_0))$ is the initial condition. The action of $K_{\rm LO}$ on the single dipole state yields the LO BK form
\begin{align}\label{eq:K1-action}
    (\bx_1\bx_2|K_{\rm LO}&= \int_{x_3}\frac{\bx_{12}^2}{\bx_{13}^2\bx_{32}^2}\, \Big[(\bx_1\bx_3\bx_2| -(\bx_1\bx_2|\Big]\,.
\end{align}
with the shorthand notation $\int_{x_i}=\int\rmd^2\bx_i/(2\pi)$.

A cross-section computed within the high energy product expansion can schematically be written as an inner product between a target operator $|S(\zeta))$ and a hard coefficient function $(H(\zeta)|$. For instance, the total DIS cross-section would read $\sigma^{\gamma^*p\to X}(x_{\rm Bj},Q^2)=(H(\zeta)|S(\zeta))$.

Inserting the completeness relation $\mathbb{I}=\sum \int_{1,...,n} |n)(n|$ where the sum runs over all possible (connected and disconnected) dipole chains $|n)$, only the first non-trivial chain contributes at the lowest order and 
\begin{align}
    \sigma^{\gamma^*p\to X}(x_{\rm Bj},Q^2)=\int_{x_1,x_2} H^{(2)}(\bx_1,\bx_2)(1-S_{12})+\mathcal{O}(\abar)\,,
\end{align}
with $H^{(2)}=(H|\bx_1\bx_2)$ the DIS LO impact factor expressed in terms of the $\gamma^*\to q\bar q$ wave-function.

\section{Scheme transformation in high energy factorization}

Besides providing a compact formulation of the Balitsky hierarchy, the main advantage of the Doi–Peliti formalism for our purposes is that it allows us to precisely define, at least in the large-$N_c$ limit, a scheme transformation in the non-linear regime. A priori, the vector $|S)$ satisfying the Balitsky hierarchy depends on the projectile rapidity variable $\zeta$ defined above. If one wishes to formulate the high-energy evolution in terms of a different rapidity variable, an additional collinear transverse scale $\mu$ must be introduced on dimensional grounds, such that the combination of $\mu$ and $\rho^+$ (or $\zeta=\rho^+ P^-$) that defines the new factorization scale has the correct dimension. For instance, high-energy factorization along the symmetric variable $k^+/k^-\sim (k^+/k_\perp)^2$ as in $\gamma^*-\gamma^*$ collision would be formulated in terms of $(\rho^+/\mu)^2$, while a factorization in terms of the target rapidity variable $P^-/k^-\sim (k^+P^-)/k_\perp^2$ would be formulated in terms of $\zeta/\mu^2$. We therefore introduce a similarity transformation generated by a $\mu$-dependent operator $L(\bar\alpha_s,\mu^2)$ in order to implement the corresponding scheme transformation:
\begin{align}
         |\bar S(\zeta,\mu^2)) =\exp\left(-\bar\alpha_s L(\bar\alpha_s,\mu^2)\right)|S(\zeta))\,,\label{eq:allorder-composite-dipole}
\end{align}
such that the vector $|\bar S(\zeta,\mu^2))$ only depends on the desired ratio of the initial rapidity factorization scale $\zeta$ and $\mu$, like $\zeta/\mu^\sigma$ where $\sigma$ is an arbitrary power to keep the discussion general at this stage. 

Enforcing that $|\bar S)$ only  depends on $\zeta/\mu^\sigma$ leads to the equation
\begin{align}
\frac{2}{\sigma}\frac{\partial \bar S}{\partial \ln \mu^2}
+ \frac{\partial \bar S}{\partial \ln \zeta} = 0 \,,
\label{eq:derivative-relation}
\end{align}
This relation constrains the operator $L$ order by order in perturbation theory. In particular, expanding $L= L_{\rm LO}+\bar\alpha_s L_{\rm NLO}+...$ in powers of $\bar\alpha_s$, eq.\,\eqref{eq:derivative-relation}, together with eq.\,\eqref{eq:allorder-composite-dipole}, implies that $\partial L_{\rm LO}/\partial \ln\mu^2=\sigma K_{\rm LO}/2$. The solution to this equation is not unique: the constant term in the polynomial in $\ln\mu^2$ remains unconstrained, reflecting the residual freedom in the choice of scheme. At order $\bar\alpha_s$, we then have
\begin{align}
        (\bx_1\bx_2|L_{\rm LO} =\frac{\sigma}{2}\int_{x_3 } \frac{\bx_{12}^2}{\bx_{13}^2\bx_{32}^2}\ln(\mu^2\br_{123}^2) \Big[ (\bx_1\bx_3\bx_2|-(\bx_1\bx_2|\Big]\,, \label{eq:L1-action}
\end{align}
where the transverse coordinate $\br_{123}$ is arbitrary. The action of $L_{\rm LO}$ on other states follows from the Leibniz rule. The scheme $\sigma=1$ and $\br_{123}=\bx_{13}\bx_{32}/\bx_{12}$ is related to the conformal dipole transformation introduced in~\cite{Balitsky:2009xg} which would be relevant for symmetric collisions where $k^+/k^-$ is the natural slicing variable. In the following, we shall study the scheme transformation to the target rapidity factorization scheme, with $\sigma=2$ and $\br_{123}$ given by $\br_{123}^2=|\bx_{13}||\bx_{32}|$, a choice which simplifies the RG evolution of $|\bar S)$. Other possible choices are $\br_{123}^2=\bx_{12}^2$ --- this corresponds to the evolution equation in $k^-$ without kinematic constraint proposed in~\cite{Ducloue:2019ezk} --- or $\bx_{123}^2=|\bx_{13}||\bx_{32}|e^{(\bx_{23}^2-\bx_{13}^2)\ln(\bx_{13}^2/\bx_{23}^2)/(2\bx_{12}^2)}$ as obtained by using a $k^-$ regulator in the one-loop correction to the DIS structure functions~\cite{Altinoluk:2025tms}.

Since the physical cross section cannot depend on the choice of scheme, the redefinition of the state vector in eq.\,\eqref{eq:allorder-composite-dipole} has to be accompanied by a compensating redefinition of the coefficient functions,
\begin{align}
(\bar H(\zeta,\mu^2)|=(H(\zeta)|\exp\left[\bar\alpha_s L(\bar\alpha_s,\mu^2)\right]\,.
\end{align}
The same transformation acts on the evolution kernel, which after expanding in powers of $\abar$ gives,
\begin{align}
\bar K & \equiv \rme^{ - \abar L }  K,  \rme^{  + \abar L}=  K  + \abar [ K,L]+\frac{\abar^2}{2!}[[K,L],L]+\mathcal{O}(\abar^3)\,,\label{eq:kernel-transformation}
\end{align}
where the second equality follows from the Campbell identity. A particularly useful consequence of this construction is that the resulting kernel $\bar K$ is independent of the transverse scale $\mu$ to all orders in $\abar$, $\rmd \bar K / \rmd \ln\mu^2 =0
$~\cite{Boussarie:2025mzh}. One sees that the scheme transformation brings an additional term to the NLO BK kernel given by $\bar\alpha [K_{\rm LO},L_{\rm LO}]$. The computation of this commutator has been done in~\cite{Boussarie:2025mzh} for $L$ given by eq.\,\eqref{eq:L1-action} with $\sigma=2$. When applied to the dipole $(\bx_1\bx_2|\bar S(Y))$ with $Y=\ln(\zeta/\mu^2)$, one gets the following modified NLO BK equation 
\begin{align}
\frac{\partial \bar{S}_{12}}{\partial Y} &=  \ \bar{\alpha} \int_{x_3} \frac{x^2_{12}}{x^2_{13}x^2_{23}} \Big[ \bar{S}_{13} \bar{S}_{23} - \bar{S}_{12} \Big] \Bigg( 1 + \bar{\alpha} \Big[ \frac{1}{12} \beta \Big\{  \ln ( x_{12} \mu_R) - \frac{x_{13}^2 - x_{23}^2}{x^2_{12}} \ln \Big( \frac{x_{13}^2}{x_{23}^2} \Big)\Big\} + const \Big] \Bigg) + \nonumber \\
    &+ \frac{1}{2} \bar{\alpha}^2 \int_{x_3,x_4} \ K_{1234} \Big[ \bar{S}_{13} \bar{S}_{34} \bar{S}_{42} - \bar{S}_{13} \bar{S}_{32} \Big] +  \frac{n_f}{2N_c}\bar{\alpha}^2 \int_{x_3,x_4} \ K_{f} \Big[ \bar{S}_{32} \bar{S}_{14} - \bar{S}_{32} \bar{S}_{13} \Big] + \nonumber \\
    &{\color{blue}  +  \frac{1}{2} \bar{\alpha}^2  \int_{x_3, x_4} \ \frac{x^2_{12}}{x^2_{13}x^2_{23}} \frac{x^2_{32}}{x^2_{34}x^2_{24}} \ln \Big(\frac{x_{34}^2}{x_{14}^2} \Big) \Big[ \bar{S}_{13} \bar{S}_{34} \bar{S}_{42} - \bar{S}_{13} \bar{S}_{32} \Big]}  \nonumber \\
    & {\color{blue}+ \frac{1}{2} \bar{\alpha}^2 \int_{x_3,x_4} \ \frac{x^2_{12}}{x^2_{13}x^2_{23}} \frac{x^2_{13}}{x^2_{14}x^2_{34}} \ln \Big(\frac{x_{34}^2}{x_{24}^2} \Big) \Big[ \bar{S}_{14} \bar{S}_{43} \bar{S}_{32} - \bar{S}_{13} \bar{S}_{32} \Big]}\,, \label{eq:nlobk-sbar}
\end{align}
where the blue terms come from the commutator. Here $\beta$ is the one-loop coefficient of the QCD $\beta$-function. The definition of the various kernels and the constant term in the first line can be found in~\cite{Boussarie:2025bpq}. As is clear from eq.\,\eqref{eq:nlobk-sbar}, the scheme transformation from the factorization in $k^+$ to the factorization in $k^-$ has canceled exactly the double collinear logarithm present in the original $k^-$-ordered NLO BK equation derived in~\cite{Balitsky:2007feb,Balitsky:2009xg}. On the other hand, the new terms in blue are free of collinear logarithmic enhancement; they only contribute in the anti-collinear regime associated with very small daughter dipoles compared to the parent ones~\cite{Boussarie:2025mzh}.

\section{Numerical analysis of the rotated NLO BK equation }

We now study the solutions to eq.\,\eqref{eq:nlobk-sbar} numerically. For the sake of simplicity, we focus on the minimal case of a fixed running coupling at the NLO order. We set its value to $\alpha_s=0.3$. The scale of the running coupling at LO appearing in the first line of eq.\,\eqref{eq:nlobk-sbar} is set in such a way as to remove the terms proportional to  $\beta$, consistent with the BLM prescription \cite{Brodsky:1982gc}. The goal of the numerical tests is to confirm that the full equation is free of collinear logarithmic enhancements, i.e., it yields a valid solution in an appropriate range of $Y$. We performed two tests using different initial conditions: one corresponding to the GBW model \cite{Golec-Biernat:1998zce} and one corresponding to the MV model \cite{McLerran:1993ka,McLerran:1994vd}.

\begin{figure}
    \begin{center}
        \includegraphics[width=0.5\textwidth, angle=270]{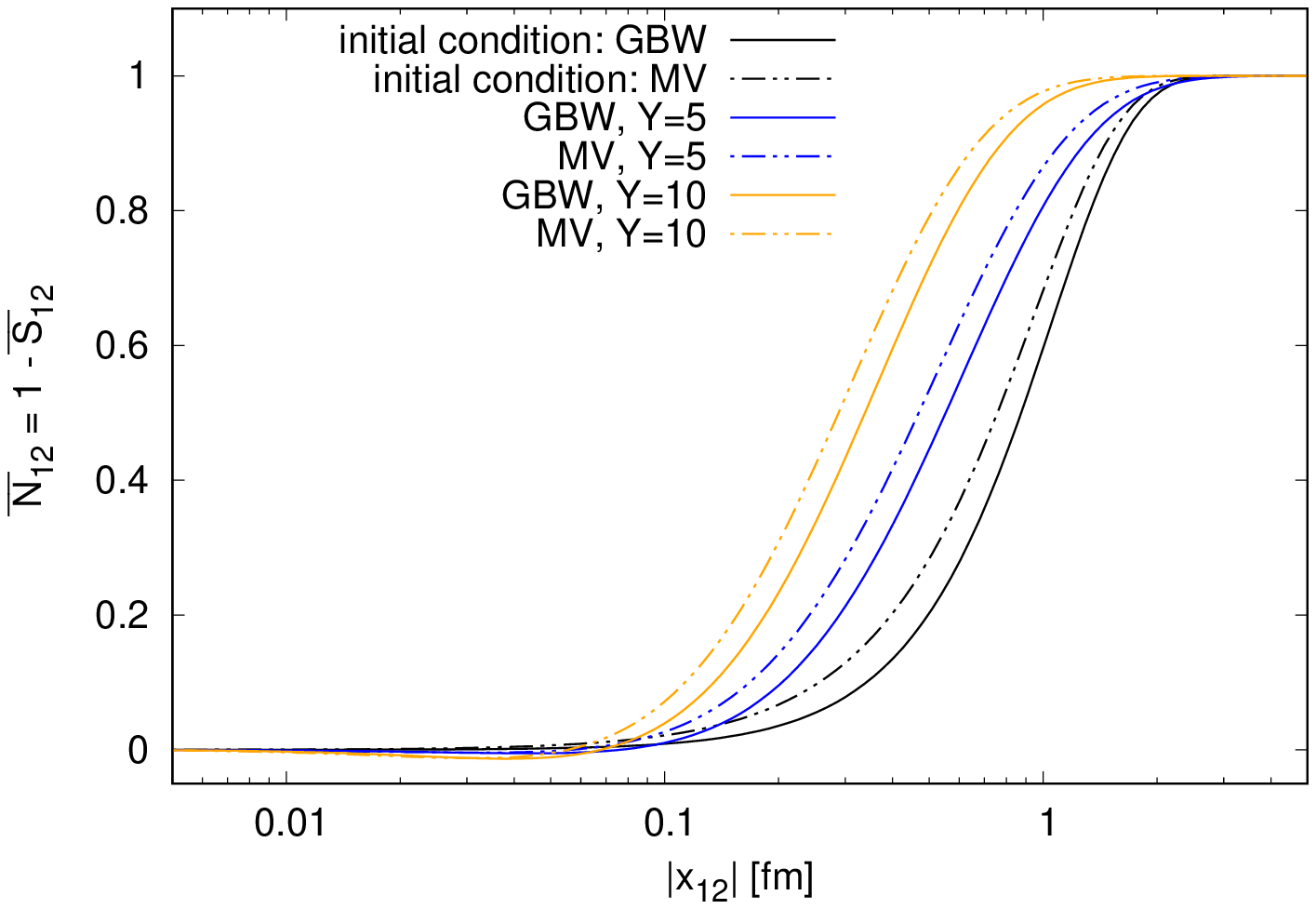}
    \end{center}
    \caption{Solutions of the full eq.\,\eqref{eq:nlobk-sbar} for different initial conditions: GBW (dashed) and MV (solid) up to rapidity $Y=10$. Both solutions remain stable under evolution. The scale of the running coupling in the first line of eq.\,\eqref{eq:nlobk-sbar} was set such that the terms proportional to $\beta$ vanish according to the BLM prescription \cite{Brodsky:1982gc}. All other coupling constants appear at the NLO order and therefore were set to a constant value of $\alpha_s=0.3$.} 
    \label{fig:results}
\end{figure}

We integrated the right-hand side of eq.\,\eqref{eq:nlobk-sbar} on a logarithmic grid for $|x_3|$ and $|x_4|$ with 256 sites spanning a range $|x| \in ( 10^{-5}, 30)$ fm. The two angles were discretized using 24 sites. All integrations have been performed using the $\frac{3}{8}$ Simpson rule. We use quadruple numerical precision for the evaluation of the integrand containing the $K_{1234}$ kernel; otherwise, we use double numerical precision for all remaining computations. In the results shown in fig.\,\ref{fig:results}, the terms proportional to $n_f$ were set to zero. The running coupling was frozen at a value of $\alpha_s=0.7$,
\begin{equation}
    1/\alpha_s(r) =b_0\log(4C^2/(r^2\Lambda_{QCD}^2)+\mu_{IR})\,,
\end{equation}
where $\mu_{IR} = \exp( 1/ (0.7 b_0))$, $C=1$, $\Lambda_{QCD}=0.241$ GeV,
and $b_0 = \frac{33-2n_f}{12 \pi}$. The evolution was calculated using the second-order Runge-Kutta method. 

We show our results in fig.\,\ref{fig:results}. We chose the initial saturation scale $Q_0=0.375$ GeV \cite{Korcyl:2026nrz}. In both cases, with the GBW initial condition (dashed) and the MV initial condition (solid), the solutions to the full equation remain stable up to $Y=10$. The onset of possible collinear single logarithmic instabilities appears at larger rapidities, visible as a small negative structure for $|x_{12}| \approx \mathcal{O}(0.05)$ fm. 

\section{Summary and outlook }

In this paper, we have presented a new theoretical framework for high-energy factorization in the nonlinear regime that allows us to treat scheme transformations consistently. Scheme ambiguity is an unavoidable feature of truncated perturbation theory and resummation, and high-energy QCD factorization is no exception to this general fact; it is therefore crucial to establish a framework that enables one to consistently move from one scheme to another. In the context of the NLO BK equation, we have defined a scheme transformation from the dipole (or projectile) scheme, in which factorization is implicit in $k^+$ (with the plus component being the large component of the dilute projectile four-momentum), to the DIS scheme, in which factorization is performed in $k^-$. This scheme transformation removes the anomalous double collinear logarithm originally present at NLO in the BK equation. We have numerically demonstrated the stability of the resulting equation with running coupling for phenomenologically relevant values of the evolution variable $Y=\ln(1/x)$. In the future, we plan to combine this stable NLO BK evolution with the “rotated” NLO impact factor, according to the scheme transformation law presented here.

\textit{Acknowledgements.} P.~C. is funded by the Agence Nationale de la Recherche under
grant ANR-25-CE31-5230 (TMD-SAT). Y.~M.~T. was supported by the U.S. Department of Energy under Contract No. DE-SC0012704. 
We are grateful for the support of the Saturated Glue (SURGE) Topical Theory Collaboration, funded by the U.S. Department of Energy, Office of Science, Office of Nuclear Physics. 
Numerical calculations were performed on the LUMI supercomputer under the time allocation: project\_465002091 (Calculating predictions for EIC physics). We gratefully acknowledge the Polish high-performance computing infrastructure PLGrid (HPC Center: ACK Cyfronet AGH) for providing computer facilities and support within the computational grant no. PLG/2024/017690. P.~K. was supported by the Polish National Science Center (NCN) grant No. 2022/46/E/ST2/00346. P.~K. thanks the EIC Theory Institute at BNL for its support and hospitality.

\bibliographystyle{apsrev4-1}

\bibliography{nlobk-references.bib}

\end{document}